# What does it mean to be "representative"?

Jacqueline E. Rudolph[1]
Yongqi Zhong[1]
Priya Duggal[1]
Shruti H. Mehta[1]
Bryan Lau[1]

[1]Department of Epidemiology, Bloomberg School of Public Health, Johns Hopkins University, Baltimore, Maryland

Correspondence:   Jacqueline E. Rudolph, PhD
                  Department of Epidemiology
                  Bloomberg School of Public Health
                  Johns Hopkins University
                  615 N Wolfe St.
                  Baltimore, MD 21205
                  (443) 287-4740
                  jacqueline.rudolph@jhu.edu

**ABSTRACT**

Medical and population health science researchers frequently make ambiguous statements about whether they believe their study sample or results are "representative" of some (implicit or explicit) target population. Here, we provide a comprehensive definition of representativeness, with the goal of capturing the different ways in which a study can be representative of a target population. We propose that a study is representative if the estimate obtained in the study sample is generalizable to the target population (either due to representative sampling, estimation of stratum specific effects, or quantitative methods to generalize or transport estimates) or the interpretation of the results is generalizable to the target population (based on fundamental scientific premises and substantive background knowledge). We explore this definition in the context of four COVID-19 studies, ranging from laboratory science to descriptive epidemiology. All statements regarding representativeness should make clear the way in which the study results generalize, the target population the results are being generalized to, and the assumptions that must hold for that generalization to be scientifically or statistically justifiable.

**INTRODUCTION**

It is common if not a requirement for medical and population health science researchers to consider the inferences from a study beyond the context of their analysis. Accordingly, many papers mention if their study sample is *representative of* or study results *generalize to* some implicit or explicit target population; others refer to a lack of generalizability or representativeness as a limitation. Despite how frequently it has been discussed and debated (1–4), in common practice, what it means to be "representative" remains ambiguous. Here, we propose a comprehensive definition of "representative" and discuss representativeness in the context of different types of studies. We presume perfect internal validity of study results; in any real-world study, bias in the study results will need to be weighed alongside sample representativeness (5).

**What is representativeness?**

The ambiguity in meaning arises in part because the word "representative" has a broader meaning in English and a more technical definition, and it is not always clear which definition is being used. In a 2013 series of commentaries on representativeness (1–4), the concept was defined as occurring when the study sample is a simple random sample of the target population, i.e., the sample that arises through representative sampling. Another definition is that the study sample and the results obtained merely resemble what would be expected in the target population (6). The first definition is more precise and implies a high standard for study design, while the second encompasses a variety of possible interpretations. There is also the related concept of external validity in causal inference, which incorporates specific quantitative methods to generalize or transport results from one sample to another and the assumptions needed to make causal statements in the target population (5,7,8).

We suggest that a study sample is representative of a well-defined target population when the results estimated in that sample are generalizable to the target population. We consider two ways study results can generalize to the target population.

A sample is representative if its results are <u>generalizable in estimate</u>. For a given estimand (e.g., risk difference, odds ratio, population mean), the *estimate* obtained in the study sample is the same within a margin of error as what would be estimated in the target population. This can be achieved if the distributions of key covariates are the same as in the target population, as would occur in expectation with random sampling. These key covariates are those that affect the variable under study (e.g., an outcome) and thus are potential effect measure modifiers of the effect of an exposure on that variable. This particular way of generalizing the *estimate* aligns with the idea of "representative sampling." More generally, if the distribution of the key covariates differs between the sample and target population, we can still generalize the *stratum-specific estimates* from the sample to the target population within strata defined by the key covariates. While this requires that all the key covariates be measured, the size of the covariate strata need not exactly match the target population. Generalizing the *estimate* obtained in one's sample might be considered the primary goal when intending to quantitatively inform policy interventions or when obtaining effect estimates in the target population is impossible or infeasible (9).

A natural extension of generalizing the stratum-specific estimates to a target population are methods to estimate the overall mean of an outcome or the average effect of an exposure on an outcome (rather than a stratum-specific estimate) in the target population by controlling for the effect measure modifiers that differ between the samples through weighting or standardizing (10). Such approaches require measuring all relevant effect measure modifiers, meeting certain identifiability conditions, and often making model specification assumptions (7,8). In such cases, the sample may not be representative as observed, but it could be made representative under these conditions.

A sample is representative if its results are <u>generalizable in interpretation</u>. While the estimates obtained in the sample are not quantitatively the same (within a margin of error) as those that would be estimated in the target population, one can hypothesize based on background knowledge that the *interpretation* and general inference around the estimate would remain the same. Any study that is representative in interpretation could theoretically be made representative in estimate if all relevant effect measure modifiers were measured and accounted for; however, that is not always possible when the study sample is distant from the target population (e.g., laboratory mice to humans). Generalizing the *interpretation* of one's results to external populations is done in most studies but might be considered the primary goal in studies examining fundamental laws of nature or asking research questions that are relatively independent of historical and environmental context. However, generalizing the interpretation should be done cautiously, as it is based on hypotheses that the mechanisms in the study sample and target population are identical.

In summary, we consider a sample to be representative of a target population if its results can be generalized to that target population either in *estimate* or in *interpretation*. Any statements made regarding the representativeness of the study need to make this further qualification. Is it the *estimate* obtained or the *interpretation* of the results that are generalizable to the target population? Researchers should also do what they can to safeguard their results from being applied incorrectly. Even in studies where there is strong scientific rationale for generalizing the *interpretation* of results to the target population, researchers may need to mention that the *estimate* obtained in the sample should not be naively generalized to the target population.

There are two points related to defining representativeness that are worth highlighting. First, irrespective of the way in which a sample is representative, the target population must be clearly defined. Stating that a sample is representative is meaningless unless one specifies what population it represents or its results are being applied to (5).

Second, one must be clear about the assumptions required for generalizing to the target population. When generalizing the *estimate*, these assumptions might be made based on one's knowledge of the relevant effect measure modifiers and the validity of the statistical models being used. If one attempted to generalize the estimate but failed to account for an important effect measure modifier that affected selection into the sample, then the assumptions would be violated, and the estimate would not be representative. When generalizing the *interpretation*, the assumptions might be made based on one's knowledge of basic scientific premises or the validity of one's animal model. If one attempted to generalize the interpretation but the scientific principles underlying that generalization did not hold (e.g., the validity of the animal model for describing human physiology), then the assumptions would be violated, and the inferences in the study would not be representative. In either case, the way to truly test whether the assumptions held would be to estimate the effect of interest in the target population. While we often generalize in estimate because this would not be feasible, we generally consider this necessary to prove hypotheses regarding generalization of interpretation, especially when the sample is highly removed from the target population (e.g., cell line vs. human population).

**STUDY EXAMPLES**

To make these concepts more concrete, we explore several example studies related to the COVID-19 pandemic.

**Laboratory Science**

At the start of the COVID-19 pandemic, there were no approved antiviral drugs for infection or disease. To assess whether antiviral drug molnupiravir was effective for treating COVID-19, in one animal model study, researchers administered molnupiravir to mice with human lung tissue before and after infection with SARS-CoV-2, using doses scaled from appropriate human levels to the mouse model (11). They found that a 2-day course of treatment, starting 24h after infection, significantly reduced SARS-CoV-2 viremia in lung tissue.

- <u>Target population?</u> All humans with recent SARS-CoV-2 infection.
- <u>Generalizability of interpretation</u>? Yes. We likely can hypothesize that the beneficial effect of molnupiravir observed in the mice would be observed in humans, based on the validity of the human lung tissue model and the observance of a similar pathological response to COVID-19 in the lung tissue of the mice as has been seen in the lung tissue of COVID patients.
- <u>Generalizability of estimate</u>? No. While the study used human lung tissue in mice and administered an appropriately scaled dose, the lung tissue was otherwise isolated from human biology, and mouse immune responses differ from those seen in humans in a manner that would be difficult if not impossible to quantify.
- <u>Big picture</u>? In this animal model study, generalizing the interpretation of the results was the primary goal. Generalization of the estimate was not relevant, as the drug would be tested further in human studies. However, underlying animal models or cell lines is the assumption that the strength of the unaccounted-for effect measure modifiers (which are likely unknown) and the difference in the distribution of the effect measure modifiers between the study sample and human target population is not large enough to change the inference being made. This strong assumption is why animal studies are followed up by clinical trials to ensure that interpretation does indeed generalize and to obtain a quantifiable estimate of the effect in humans.

**Randomized Control Trial**

The efficacy of molnupiravir for treatment of COVID-19 was investigated in a Phase 3 clinical trial in non-hospitalized, unvaccinated adults with mild to moderate COVID-19 disease (12).

Trial researchers reported that the 29-day risk of hospitalization or death among participants randomized to molnupiravir was 6.8%, compared to 9.7% of participants randomized to placebo. They concluded that treatment with molnupiravir within 5 days of infection reduced the risk of hospitalization or death.

- Target population? All non-hospitalized adults with recent SARS-CoV-2 infection.
- Generalizability of interpretation? Yes. It would be reasonable to hypothesize that molnupiravir would have a beneficial impact if administered to those infected with SARS-CoV-2, even beyond this sample of non-hospitalized participants with moderate illness. The researchers noted that the trial participants were similar to real-world patients in terms of having 1 or more risk factors for severe illness. We would base this hypothesis on our understanding of the drug's biological mechanism and the validity of a properly conducted double-blind, placebo-controlled trial.
- Generalizability of estimate? We would not be able to generalize the trial's estimate to the entire target population, even if we could control for post-randomization factors such as non-compliance. This is because the target population includes all non-hospitalized individuals recently infected with SARS-CoV-2, not just those who were unvaccinated. It is reasonable to assume that vaccination status is an effect measure modifier of the effect of molnupiravir on hospitalization, and the trial sample did not include vaccinated participants. We could generalize within the unvaccinated stratum, provided all other effect measure modifiers were measured.
- Big picture? In this randomized control trial, generalizing the interpretation regarding drug efficacy to the target population was the primary goal. If it was possible, generalizing the estimate to the target population would be useful for predicting how molnupiravir would perform in practice but may not be immediately required for the study results to be meaningful. Further studies would likely need to be conducted to generalize the interpretation to other target populations, like children recently infected with SARS-CoV2.

**Observational Study**

Researchers used testing, hospital, and vaccine registry databases to build an observational cohort of vaccinated adults living in New York, with age-matched unvaccinated controls (13). Their goal was to assess the effectiveness of COVID-19 vaccines for preventing infection with SARS-CoV-2 and hospitalization due to COVID-19 in the general population. They found that vaccine effectiveness for preventing infection was highest in the week of May 1, 2021 (93.4%), when prevalence of the Delta variant was negligible, but that effectiveness declined as Delta became more prevalent – to a low of 73.5% the week of July 10, 2021. In contrast, the effectiveness for preventing hospitalization did not wane during this same calendar period. The researchers concluded that their findings regarding decreased vaccine effectiveness over time was evidence in support of booster vaccines.

- Target population? Adult residents of New York State.
- Generalizability of interpretation? Yes. There is little reason to suspect that vaccines would not be effective against COVID-19 and hospitalization or that we would observe different trends in vaccine effectiveness over time among New York residents who were not included in this study (or residents of other states).
- Generalizability of estimate? With additional data, yes. The registry-based study included a wide range of ages and the different vaccine types (Pfizer, Moderna, and Johnson & Johnson). The paper did not report the sex and race distributions of their cohort, but, with data on these potential effect measure modifiers, one would likely be able to generalize the estimate from this study to the broader New York population.

- Big picture? Here, generalizing the interpretation and the estimate are both important. Generalizing the estimate to the target population requires more effort both from those designing the study and from those analyzing the data but could be incredibly useful for informing COVID-19 prevention efforts in New York. If we wished to generalize to target populations beyond New York (e.g., the entire US), we would need to make assumptions about whether there are effect measure modifiers that differ between New York and the entire US and whether we have them measured.

**Descriptive Study**

Researchers sought to capture the burden of COVID-19 among people who inject drugs in the San Diego-Tijuana area (14). Participants from both cities were recruited using street outreach and mobile vans. Blood samples and nasal swabs were collected to test for the presence of SARS-CoV-2 antibodies and RNA. None of the 485 participants had detectable SARS-CoV-2 RNA, but 140 (36.3%) were seropositive based on the presence of antibodies. This was larger than the prevalence reported in the general population for either city. There were no trends in prevalence of antibodies to SARS-CoV-2 over the study period (October 2020-June 2021).

- Target population? People who inject drugs in the San Diego-Tijuana region.
- Generalizability of interpretation? Yes. It is likely the case that people who inject drugs in this region have a higher prevalence of SARS-CoV-2 than the general population, even beyond the time frame and sample studied.
- Generalizability of estimate? Perhaps. The 36.3% prevalence of SARS-CoV-2 could potentially be generalized to the full sample of people who inject drugs in this region, at least during the time frame examined. We would be unable to generalize the estimate to other points in the pandemic, with different SARS-CoV-2 strains and levels of community exchange.

- Big picture? Just as in the observational study, generalizing both the estimate and the interpretation are important for assessing the relevance of this study. We note here that the target population selected was significantly narrower than those of the previous studies, but this reflects the research and public health goals of the study. The researchers likely could not make statements regarding representativeness to broader target populations (e.g., all people who inject drugs in the US) without further evidence.

**CONCLUSIONS**

We have established the idea that a study sample can be representative of a target population if one of the following is true: (1) the estimate obtained in the study sample is generalizable to the target population or (2) the interpretation of the study results is generalizable to the target population. Whether a study sample can be representative of a target population depends on the effect measure modifiers or more generally, the variables affecting the outcome. Capturing the effect measure modifiers in the study will allow for the study's estimate to represent the target population. Even in the absence of capturing all relevant effect measure modifiers, we can often allow for the study to be representative in terms of interpretation and inference, as this requires a less stringent assumption than generalizing the study estimate. The examples provided give guidance on how one might determine whether the study sample from different types of research is representative and whether, for the specific research question, generalizing the estimate or the interpretation was the priority.

One question is whether generalizing the interpretation or the estimate is intrinsically more important for health research and for science broadly. Some argue that generalizing the interpretation is the primary aim of scientific inference and thus should be our goal in most studies (1). The underlying premise is that the goal of science is the discovery of universal knowledge about nature that will hold true in most places and times. If we view health research from this lens, then generalizing the interpretation is what matters.

In contrast, generalization of the estimate can never be universal. It will always be tied to a specific scientific or public health question, the target estimand, and the estimate obtained and vary based on the distribution of key effect measure modifiers across time and populations. This is exactly why it is critical in the applied setting. To inform policies and interventions, we must be able to predict health outcomes in human populations beyond those we studied. One could take the argument even further and state that these endeavors of statistical inference are just as informative for science as the inferences above. Science can be about discovering laws of nature; it can also seek to understand particular facets of nature. For some areas of health research, such as epidemiology and other population health sciences, the facet of nature under study is disease as it occurs in humans at a population level, and disease as it occurs in populations cannot (and perhaps should not) be fully divorced from time, place, history, and social context.

However, our comprehensive definition of representativeness does not treat either generalization of estimate or interpretation as more relevant, as that depends on the research question and study design. Health researchers both develop the universal knowledge related to the health of populations and investigate how that knowledge can be applied to improve the health of populations, and the two ends of the research spectrum are fundamentally linked. What is important, then, is that researchers be clear on the manner in which their results can be reasonably generalized when they say their study is "representative" and the assumptions underlying that statement.

**ACKNOWLEDGEMENTS**

None.

**FUNDING**

This work was supported in part by National Institutes of Health grants R01-CA250851 and U01-DA036297.

**COMPETING INTERESTS**

All authors have completed the ICMJE uniform disclosure form at http://www.icmje.org/disclosure-of-interest/ and declare: the corresponding author had financial support from National Institutes of Health grants R01-CA250851 and U01-DA036297 for the submitted work; all authors declare no financial relationships with any organisations that might have an interest in the submitted work in the previous three years; no other relationships or activities that could appear to have influenced the submitted work.